\documentclass{phyeauth}
\usepackage{graphicx}
\usepackage{amsmath}
\usepackage{amssymb}

\begin{document}

\begin{frontmatter}

\title{
Model for Breakdown of Laminar Flow of a Quantum Hall Fluid Around
a Charged Impurity: Comparison with Experiment}

\author{A. M. Martin},
\author{K.A. Benedict},
\author{F.W. Sheard}
and
\author{L. Eaves}

\address{School of Physics and Astronomy, University of Nottingham, Nottingham, NG7 2RD, UK}

\begin{abstract}
In samples used to maintain the US resistance standard the
breakdown of the dissipationless integer quantum Hall effect
occurs as a series of dissipative voltage steps. A mechanism for
this type of breakdown is proposed, based on the generation of
magneto-excitons when the quantum Hall fluid flows past an ionised
impurity above a critical velocity. The calculated generation rate
 gives a voltage step height in good agreement
with measurements. We also compare this model to a hydrodynamic
description of breakdown.
\end{abstract}

\begin{keyword}
Quantum Hall Effect \sep breakdown \sep magnetoexciton \sep
hydrodynamics
\PACS 73.43.-f \sep 73.43.Cd \sep 73.43.Lp \sep 47.37.+q
\end{keyword}
\end{frontmatter}

In the integer quantum Hall effect (QHE) regime, a two-dimensional
electron fluid carries an almost dissipationless current and the
ratio of the current, to the Hall voltage is quantized in units of
$e^2/h$. However, above a critical current, the dissipative
voltage, $V_x$, measured  along the direction of current flow,
increases rapidly, leading to breakdown of the QHE. For certain
samples, including those used to maintain the US resistance
standard at the National Institute of Standards and Technology
(NIST) \cite{Cage,Lavine}, breakdown occurs as a series of steps,
in $V_x$, of height $\Delta V_x \approx 5mV$. Below we examine
this step height using two distinct methods. First we perform a
fully quantum calculation for the rate of formation of e-h pairs
due to a single charged impurity, as a function of electric field,
$E_y$. In doing this we obtain a relationship between $V_x$ and
$E_y$. We find that as $E_y$ increases $V_x$ increases rapidly to
a maximum and then falls slowly. We then argue that the only
stable points on this graph are firstly when $V_x$ is less than
the small dissipative voltage governed by $\sigma_{xx}$ and
secondly when $V_x$ is a maximum. Hence we are able to estimate
the step height $\Delta V_x$. We then consider a hydrodynamical
model \cite{Eaves} of this process starting from an effective
Euler equation. By examining small pertubations in the velocity
field we find a direct analogy to the elastic inter-Landau level
tunneling condition \cite{Fred}, from which we can again estimate
$\Delta V_x$. Finally, we compare the results of the two
approaches with experiment and comment on how this work relates to
bootstrap electron heating (BSEH) \cite{Komiyama} model of the
breakdown of the integer QHE.

The starting point is the Fermi golden rule, which we use to
calculate the rate of generation of e-h pairs, $W$, due to a
single charged impurity in a uniform electric field. Using the
Landau gauge, this gives
\begin{figure}[h]
\begin{center}\leavevmode
\includegraphics[width=0.55\linewidth]{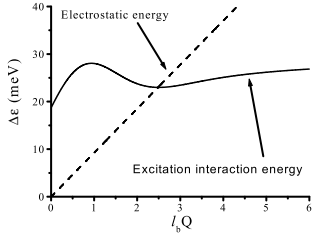}
\caption{The magneto-exciton mode energy, $\hbar \omega_c+e^2/(4
\pi \kappa l_B) \Delta_{0,(1)}[Q]$ (solid line) for excitations
from $n=0$ to $n=1$, for $E_y=0$; the dashed line is the
electrostatic energy, $eE_yl_B^2Q$. $B=12.3T$ and $E_y=1.5 \times
10^{6}V/m$.}\label{figurename}\end{center}\end{figure}
\begin{eqnarray}
& &W=4\sum_{k_x,k_x^{\prime}}W_{n,(n+1)}= 4\sum_{k_x,k_x^{\prime}}
\frac{2 \pi}{\hbar} \delta(\epsilon_{n}-\epsilon_{(n+1)}) \times
\nonumber
\\ & &  \left| \int_{r} d^3r
\left|\Xi_0(z)\right|^2\phi_{n}(r_{\perp},k_x)\phi_{(n+1)}(r_{\perp},k_x^{\prime})V(r)\right|^2
,
\end{eqnarray}
\noindent where $V(r)$ is the impurity Coulomb potential,
$\phi_{n}(r_{\perp},k_x)$ is the electronic eigenfunction in the
$x$$y$ plane and $\Xi_0(z)$ is the envelope function of the first
electronic sub-band. We assume that the $n$ lower Landau levels
are filled and the ($n+1$) level is empty. Hence there is no
static screening of the impurity charge. Calculating the
transition rate out of state ($n,k_x$), by summing over
$k_x^{\prime}$ in Eq. (1) we obtain the following energy
conservation condition
\begin{equation}
\Delta \epsilon = \hbar \omega_c+\frac{e^2}{4 \pi \kappa l_B}
\Delta_{n,(n+1)}[k_x^{\prime}-k_x]-eE_yl_B^2(k_x^{\prime}-k_x)=0,
\end{equation}
where $\omega_c$ is the cyclotron frequency, $l_b$ is the magnetic
length, $\kappa$ is the dielectric constant, $E_y$ is the $y$
component of the Hall electric field,
$\hbar(k_x^{\prime}-k_x)=\hbar Q$ is the momentum change and
$\Delta_{n,(n+1)}[Q]$ includes the local and exchange field
corrections \cite{Lerner} for an excitation from ($n,k_x$) to
($n+1,k_x^{\prime}$). The condition given by Eq. (2) is more
precise than that presented in Ref. \cite{Fred} for elastic
inter-Landau level tunneling. The improvement comes from inclusion
of local and exchange field corrections for the excitation
process, i.e. in Ref. \cite{Fred} $\Delta_{n,(n+1)}[Q]=0$. When
the dashed line crosses the solid curve, in Fig. 1, the condition
given by Eq. (2) is met. At this point, energy is conserved in
moving an electron from the filled lower Landau level (LL) to the
unoccupied upper LL. For a given electric field, $E_y$, we then
use Eq. (1) to calculate the rate of generation of e-h pairs due
to the presence of an impurity.
\begin{figure}[h]
\begin{center}\leavevmode
\includegraphics[width=0.5\linewidth]{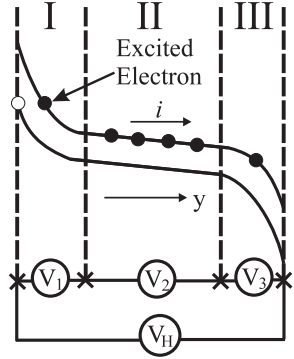}
\caption{A schematic diagram showing the variation of the Landau
level energy across the Hall bar. The e-h pairs are formed in
region I by scattering from a charged impurity at a rate $W$. The
cross-section of the bar is split into three regions. In regions I
and III, near the edges of the bar, the voltage drop is high
compared to region II in the centre of the hall bar.
}\label{figurename}\end{center}\end{figure}

 Consider a local
region of the sample where $E_y$ is large enough to create e-h
pairs at a given rate, due to scattering from a charged impurity.
A pair created close to an impurity drifts along the Hall bar at a
velocity $E_y /B$, so for a fixed generation rate, a stream of e-h
pairs moves along the Hall bar. Then, for a filling factor of 2 we
have a situation in which a small fraction of electrons in the
lower LL ($n=0$) have been replaced by holes and the previously
empty upper LL ($n=1$) contains some electrons. As the e-h pairs
move away from the high field region, the spacing between the
electron and hole in a pair increases and most pairs eventually
ionise by acoustic phonon emission. Due to the absence of empty
states into which the excited electron can relax and neglecting
weak, second order Auger processes, we can assume that all the
generated e-h pairs eventually ionise, leading to a dissipative
current $i=eW$, where $W$ is calculated from Eq. (1). This process
is shown schematically  in Fig. 2. From this figure we see that an
e-h pair
 created close to the left (hot) edge of the
Hall bar leaves a hole which ionises at the hot edge of the bar
whilst the excited electron, in the $n+1$ Landau level, relaxes
its energy due to phonon emission as it passes through regions I,
II and III. At filling factor $\nu=2$, this process gives rise to
a dissipative voltage
\begin{equation}
V_{x}= \left(\frac{hW}{2e}\right).
\end{equation}

The NIST experiments carried out at $\nu=2$ and $B=12.3T$ show a
series of up to 20 dissipative steps in $V_x$ of regular height
$\Delta V_x \approx 5mV$. In Fig. 3 we plot our calculated
dissipative voltage versus the background electric field $E_y$.
Since $W$ is strongly influenced by the overlap between the
wave-functions in the occupied ($n=0$) and unoccupied ($n=1$) LLs,
it increases rapidly at a critical electric field. This occurs
when $V_x$ is comparable to the small background dissipative
voltage governed by $\sigma_{xx}$. The rate, W, and hence $V_x$,
then reaches a maximum when the electric field is such that the
e-h pairs are formed close to the roton minimum of the
magneto-exciton dispersion curve. For a given $E_y$, for which $W$
is finite, the pairs formed at the breakdown point relax, so that
electrons in the upper LL moving towards one side of the Hall bar,
whilst holes move in the opposite direction. The presence of these
carriers screens the Hall field over much of the Hall bar. Since
the Hall voltage in the NIST experiments remains constant at its
quantized value over the magnetic field range where the
dissipative steps occur, this screening effect tends to enhance
the electric field at the breakdown point. Thus, as the critical
electric field is reached, the generation rate at the breakdown
point increases rapidly, inducing a further increase in $E_y$ due
to the screening of the Hall field in other regions of the Hall
bar. We can see how this happens in Fig. 2. The electron is
excited in region I and then decays due to phonon emission to
region II. The voltage drop, $V_2$, is small so the electron takes
a relatively long time to traverse this region. Hence, if the
excited electrons in region I are generated at a constant rate,
$W$, they tend to accumulate in region II. Since $V_H$ remains
constant in the magnetic field range over which the steps are
observed \cite{Lavine} and the accumulated electrons in region II
screen the local electric field in this region the voltage drop,
$V_2$, must reduce and hence the voltage dropped in regions I and
III must increase.  As a result, the electric field in regions I
and III increases. This increase in $E_y$ increases $W$ until it
reaches a maximum. Thus for breakdown at a single charged impurity
the system should switch between two stable states, corresponding
to $V_x=0$ and $V_x=5.6mV$, which corresponds to the maximum value
of $V_x$ in Fig. 3. This is in good agreement with NIST
experiments. In the experiments a series of up to twenty steps are
observed and we attribute each step to the formation of separate
streams of e-h pairs generated by different impurities.
\begin{figure}[h]
\begin{center}\leavevmode
\includegraphics[width=0.55\linewidth]{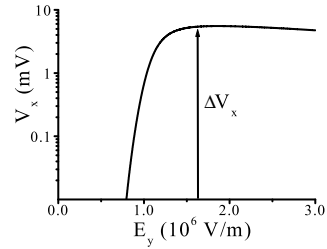}
\caption{The dissipative voltage, for an electron gas sample,
calculated from Eq. (3) as a function of $E_y$. The parameters
used refer to the experimental conditions of
[2].}\label{figurename}\end{center}\end{figure}

We now examine more closely an analogy recently proposed by one of
us \cite{Eaves} between the breakdown of the QHE and the breakdown
of laminar fluid flow around an obstacle. Our starting point for
such a model is an effective Euler equation, derived by Stone
\cite{Stone},
\begin{equation}
m^{\star}\left[{\bf \dot{v}}-\left[{\bf v} \times {\bf
\Omega}\right]\right]=e\left({\bf E}+{\bf v} \times {\bf
B}\right)-{\bf \bigtriangledown}
\left(\frac{m^{\star}}{2}\left|{\bf v}\right|^2+\mu\right),
\end{equation}
where ${\bf v}$ is the velocity field of the QHF, $\mu$ is the
local chemical potential containing all the interaction terms and
${\bf \Omega }$ is the fluid vorticity. Examining small
pertubations in the velocity and combining Eq. (4) with the
continuity equation for the density of the QHF we find, in the
limit ${\bf \bigtriangledown} \mu=0$,
\begin{equation}
\hbar \omega = \Delta \epsilon=-eE_yl_B(l_BQ)+\hbar \omega_c.
\end{equation}
For $\Delta \epsilon=0$, this result is equivalent to Eq. (2) in
the absence of interactions and corresponds exactly to the elastic
inter-Landau level tunneling condition for the breakdown of the
integer QHE \cite{Fred}. Alternatively within this hydrodynamic
model it corresponds to the condition required to generate a
vortex-antivortex pairs at zero energy. From Eq. (5), we can
deduce that for a fluid described by Eq. (4) laminar flow becomes
unstable when
\begin{equation}
E_y=\frac{\hbar \omega_c}{el_B(l_BQ)}.
\end{equation}

To make a direct comparison with our earlier quantum mechanical
calculation we need to evaluate the dissipative voltage drop along
the sample due to the generation of these vortex-antivortex pairs
from a single impurity. For a specific system we would have to
rely on a numerical simulation of Eq. (4). However, we can make
considerable progress by considering vortex shedding in classical
fluid mechanics \cite{Massey}. Consider an obstacle in the path of
a fluid. At a low fluid velocity the flow around the obstacle is
laminar. When the flow rate is increased, vortex-antivortex pairs
are formed in the vicinity of the obstacle. However, a vortex
street is not formed until the flow is fast enough to free the
vortex-antivortex pairs from the local flow field near the
obstacle. In this steady state, each vortex-antivortex pair moves
away from the obstacle at a velocity which is governed by the
background fluid velocity. This analogy suggests that the
vortex-antivortex pair in a QHF moves away from the impurity at a
velocity given by $E_y/B$. From classical hydrodynamics
\cite{Massey} it is also known that the distance between each
vortex-antivortex ($l$) pair generated is approximately three
times the separation between a single vortex and antivortex ($d$)
($d/l=0.28$). Now consider two states for our fluid: firstly, the
state where the fluid flow is laminar; then we know that the
generation rate of vortex-antivortex pairs is zero. When $\Delta
\epsilon=0$ it cost no energy to generate excitations and the flow
ceases to become laminar due to the formation of vortex-antivortex
pairs. Hence from Eq. (6) we can evaluate $E_y/B$. Dividing this
velocity by the distance between each vortex-antivortex pair
($d=0.28 l_B(l_BQ)$), to obtain the rate of generation of
vortex-antivortex pairs, we obtain the voltage drop along the Hall
bar as
\begin{equation}
V_x=\frac{0.28 \pi \hbar \omega_c}{ e (l_BQ)^2}.
\end{equation}

To estimate the dissipative voltage step height from this
hydrodynamic model we have to evaluate Eq. (7). We can do this by
referring back to the previous quantum mechanical calculation to
give us an estimate of $Ql_B$. Taking the value for $Ql_B=1.9$ for
which the quantum calculations give the maximum value for $V_x$ we
find, using Eq. (9) that $\Delta V_x=4.8mV$ for the NIST
experiment \cite{Lavine}.

In conclusion we have examined two models for the step-like
breakdown of the integer QHE. The first is a full quantum
calculation of the dissipative voltage drop along the Hall bar
generated by scattering from a charged impurity. The second method
uses an effective fluid model to describe the QHF from which we
derive the elastic inter-Landau level tunneling condition
\cite{Fred}. From this condition and drawing analogies with
classical fluid dynamics we also obtain $\Delta V_x$. In both
cases we compared our theory to the experiments of Lavine {\it et
al.} \cite{Lavine} which were carried out on the U.S. resistance
standard sample. In each case the agreement between theory and
experiment is quite good. It should also be noted that we have
also made comparisons of our models with the step-like breakdown
observed for a QHF of holes \cite{Oleg}; again our theoretical
results match experiments quite well \cite{Andy}.

The models which we have presented above are for a specific type
of breakdown in which $V_x$ increases as a series of steps of
regular height. However, many samples do not exhibit such
behaviour at breakdown. In such samples a single large increase in
$V_x$ is observed at breakdown \cite{Komiyama}. This behaviour is
generally explained by the BSEH model \cite{Komiyama}, which
describes an avalanche process which excites electrons over an
extensive area of the sample. The model presented above does not
contradict this model; in fact it complements it. Our model which
describes a
 {\it local} breakdown due to scattering from charged
impurities, can be a precurser to this avalanche breakdown which
spreads over much of the sample, i.e. in some samples it is
possible for the conditions for local charged impurity-induced
breakdown to be met before the avalanche process takes over. The
natural question then becomes what are these conditions and when
can they be met. The condition for local charged impurity induced
breakdown is sample specific; we require a locally {\it high}
electric field in the region of a charged impurity to generate a
stream of e-h pairs as compared to the total electric field across
the sample reaching a critical value to trigger BSEH. We suggest
that in the NIST samples some mechanism is inhibiting the
avalanche, thus allowing us to observe the weaker local step-like
breakdown.

This work is supported by the EPSRC.

\end{document}